\documentstyle{mn}

\begin{document}
\title {Waiting for the Big One: A New Class of SGR
Outbursts?}
\author
{\\ David Eichler \\ Department of  Physics \\ Ben-Gurion
University \\ Beer-Sheva, 84105, Israel \\
 eichler@bgumail.bgu.ac.il}
\date{Accepted xx. Received xx}
\pubyear{1995}

\maketitle
\begin{abstract}

It is suggested that magnetars eventually become unstable to a
dynamic overturning instability that destroys most of their dipole
moment in a single event. It is further suggested that such an
instability would produce a supergiant version of the largest soft
gamma ray bursts that have thus far been observed, and that  they
could be observed out to tens or even $10^2$ Mpc. Persistent
emission and afterglows from the fireball might confirm positive
identification with nearby galaxies and perhaps provide other
distinguishing features as well.
\end{abstract}

 Soft Gamma Ray (SGR) repeaters and anomalous X-ray
pulsars are now believed to be "magnetars" - neutron stars with
surface field of order $10^{15}$ Gauss or more as originally
proposed by Thompson and Duncan (1992, 1995, 1996). The energy
that drives the very large ($\geq 10^{44}$ ergs) giant flares such
as the March 5, 1979 event from SNR N49 and the Aug. 27, 1998
event from 1900+14  is apparently attributable to a sudden large
scale rearrangement of the magnetic field which releases magnetic
energy,
 while the smaller repeating bursts ($E\leq 10^{41}$ergs) seem to
be well explained as being driven by crustquakes that are
presumably the episodic yielding of the crust to magnetic strain.
As further argued by these authors, the required magnetic energy
demands that a large volume of field be released or annihilated -
many cubic kilometers. Because this is more than the thickness of
the crust cubed, the implied crustal motion is likely to be
horizontal. This would be expected given the strong stability
provided by the mass stratification in the crust.

That magnetars release their magnetic energy over the intermediate
timescale $10^4$ years - much longer than dynamical, much shorter
than Ohmic -  has been explained  (Goldreich and Reissenegger,
1992, Thompson and Duncan, 1996) as being due to relaxation of the
field in the core. Such relaxation is in general resisted by
pressure equilibrium given the alteration of  the density of the
charged fluid (i.e. electrons and protons) that the relaxation
would cause. Thus the relaxation can proceed indefinitely only
over the timescale of beta equilibration, which allows electron
and proton chemical potentials to equilibrate across different
field lines via the neutrons. The rate of beta equilibration is a
strongly increasing function of the field strength  and is of
order 10$^4$ years for field strengths of 10$^{15}$ Gauss
(Thompson and Duncan, 1996). (For $10^{12}$ Gauss fields, the
timescale would be longer than that for Ohmic dissipation.) In the
SGR's and AXP's, the light curves are typically complex enough to
indicate significant higher-order multipole components. An
internal field that is  in a complex, unrelaxed state may in fact
be a necessary prerequisite for crustal rearrangement and magnetic
activity. The  relaxation could continue to completion after the
SGR phase.

As there have been only two giant flares observed thus far from
SGR's, the "luminosity function"  is not well known, and the
details of the magnetically induced crustal rearrangement are hard
to pin down. The best one can say is that, as two SGR's have
displayed giant flares within an observation period of order 20
years,  they probably repeat about $10^2$ to $10^3$  times over
the $\sim10^4$ year ages of the objects. For each event,
presumably, only a small fraction of the magnetic energy is used
up, and may represent the simplification of the magnetic field
pattern, consistent with the observed simplification of the X-ray
pulse profiles. On the other hand, if the magnetic field is so
capable of pushing the crust around, the possibility exists that
its energy could be released in fewer, more dramatic events. The
extreme possibility is that  the entire dipole moment could be
destroyed in a single event.  (The initial spin-down of a magnetar
is a model for a cosmologically distant gamma ray burst [Usov
1992].)

    The idea that a dipole magnetic field in a star could destroy itself by
a dynamical instability was first proposed by Flowers and Ruderman
(1977). They suggested that one hemisphere would flip by 180
degrees relative to the other, in the same manner that  two
unrestrained aligned bar magnets will flip relative to each other.
This is indeed shown formally to be the case  (Ray 1980, Eichler
1982) for a simplified cylindrical geometry in a crust free case.
Wang and Eichler (1988) showed that if there is a rigid crust, the
dynamical instability becomes a secular instability, growing over
the Ohmic dissipation timescale of the crust (at least $\sim10^6$
years). Although the destruction of dipole moments was conjectured
to have possible observational consequences, e.g. in regard to the
absence of any evidence for dipole fields in accreting low mass
X-ray binary neutron stars, it was hardly supposed that the
process itself - which in the case of "classical" $10^{12}$ Gauss
fields releases only about $10^{41}$ ergs  over millions of years
deep in the neutron star's crusts - could be directly observable.
Moreover, the overturning instability demands a high degree of
symmetry in the magnetic field.  There must be a plane through
which magnetic field lines do not cross, so that the relative
rotation of the two regions separated by this plane  does not
shear the field. There was no good reason to believe that such
symmetry would ever be achieved in standard neutron star before
Ohmic dissipation in any case destroyed most of the field.

This letter suggests that in the case of magnetars  (Duncan and
Thompson, 1992, Thompson and Duncan, 1995, 1996)  there may be
reason to believe that the overturning instability a) takes place
eventually  and b) proceeds on a dynamical timescale and could be
directly observable. Magnetars, it is argued below,  teach us
several lessons that suggest why the magnetic field might achieve
a sufficiently high degree of symmetry in the case of magnetars,
though not necessarily in the case of more weakly magnetized
neutron stars, and  why the instability might be more observable
in  magnetars than in weakly magnetic neutron stars.

 Why would such an instability
be more likely than in standard neutron stars and magnetic white
dwarfs, where dipolar fields seem to persist? 1) Magnetars differ
from the other magnetized compact objects in that their magnetic
field energy dominates rotational energy. In other compact
objects, any significant differential rotation could easily
sustain a toroidal field, which prevents overturning. 2) Even in
the absence of differential rotation, the effective conservation
of plasma mass on closed flux tubes (imposed by $\beta$ stability
and magnetic flux freezing) can inhibit symmetrization of the
field. Consider, for example, a single poloidal loop of magnetic
flux that is concentrated at a particular longitude and anchored
at its lowest point  at a particular latitude by convectively
stable matter. Magnetic energy is lowered by smearing the loop
into an axisymmetric torus, still anchored at the same depth.
This, however, would greatly increase its volume and, due to the
lower plasma pressure inside the loop, would then  cost PdV work
unless the plasma passes across field lines. The tendency for
initially non-uniform magnetic flux loops to distribute themselves
more uniformly around the axis of the star, allowed by
$\beta$-equilibration in the strong magnetic field, may in fact be
the reason for giant flares in magnetars.

 In magnetars, unlike in other neutron stars,  the magnetic field is
strong enough to plastically deform or even break the crust unless
the field is very nearly axisymmetric. (It is therefore
conceivable that in some cases it could take place soon after the
formation of the field.) Thus, if the instability took place, it
could proceed on a dynamical timescale once it developed
sufficiently. Over several growth times, the overturning
instability could develop a significant amplitude even if it
originates from a highly axisymmetric field. As it proceeds, the
magnetic stresses on the crust would increase, and, for magnetar
field strengths, could eventually break or melt the crust. Beyond
this point, the instability proceeds dynamically and much of the
dipole energy of the field could be released via magnetospheric
currents, as invoked by Thompson and Duncan to explain emission of
SGR's and AXP's.

Such an event, it is suggested here,  would appear as a supergiant
version of giant SGR outbursts (hereafter referred to as SGSGRB,
for supergiant soft gamma ray bursts. Though we do not know for
certain that their spectra would be soft, they are {\it
supergiant} in the context of {\it soft} gamma ray repeaters):
Rather than the dipole field energy being released in $\sim10^2$
or more events, as seems to be the case for SGR's (which have ages
of $\sim10^4$ years and repetition times for giant events of order
tens of years), this field energy would be released in a single
event, and could be more than $\sim10^2$ times as energetic as the
giant SGR flares of March 5, 1979 and Aug. 27, 1998. The much
larger energy would be expressed both as larger magnetospheric
currents, due to crustal displacements of order the radius of the
star, and a larger area over which the surface field is sheared.

On the other hand, a SGSGRB would occur only once or twice per NS
- one time as the dipole moment is destroyed and perhaps a second
time as the quadrupole moment is destroyed (though, arguably, both
could happen in the same event). Such  bursts should thus be at
least $10^2$ times less frequent than giant flares from magnetars,
but could be detectable out to perhaps 10 times the distance, and
could be observed an order of magnitude more frequently than
extragalactic versions of the giant flares from SGR's. The Aug. 27
giant flare of SGR 1900+14 had a peak luminosity of nearly
$10^{45}$ erg/s, and could thus be detectable out to about 10 Mpc
with BATSE technology. By contrast, SGSGRB's, if they occur, might
be observable out to 100 Mpc.  They would appear to be more or
less distributed as galaxies at a fluence of $10^{-7}$ erg /cm$^2$
with a $<V/V_{max}>$ of about 0.5.

 SGSGRB's
would likely have timescales resembling those of the giant SGR
outbursts, though the following consideration suggests that there
could be some differences: The ratio of total energy to that in
contained  magnetospheric plasma is likely to be higher. Thus, the
increase in absolute luminosity that is predicted would apply only
to the beginning of the burst. The prolonged emission, which
persists over several rotation periods for the two giant flares,
and which presumably is from contained magnetospheric plasma,
would not necessarily be significantly enhanced in the SGSGRB
version and would go mostly undetected for nearly all of them.
Rather, the observed parts of most SGSGRB's would have light
curves that resemble the "tips" of the giant flare light curves.
The characteristic timescale of several tenths of seconds is what
is expected for global magnetic field-induced motion. Similarly,
one might expect the SGSGRB spectra to be soft, like the two giant
flares, but it is not claimed here that this is certain.

In any case, it has been reported that both soft bursts, and short
bursts (i.e. all but long, hard bursts) have a V/V$_{max}$
distribution that is consistent with 0.5 (e.g. Tavani 1998 and
references therein). Thus, these data admit a sizable fraction
(perhaps even more than suggested here) of GRB's to be local
unless they are both long and hard.

The rate of SGSGRB's could be as high as the rate of magnetar
production, $\sim 10^{-3}$ per galaxy-year, as we know of no
reason to expect strong beaming of the prompt emission, but this
number is somewhat uncertain. If they are detectable from the
$3\times 10^4$ to $10^5$ nearest galaxies, then, in principle up
to 15 to 50 per year could be observed with a $2\pi$ solid angle
detector. While there is some chance that SGSGRB's could account
for some major subclass of GRB's  (e.g.  all short bursts) given
the uncertainties, we believe it is also worth considering that
they represent a smaller subclass of GRB's.

Pinpointing a sufficient number of SGSGRB's in their host galaxies
would confirm the physical reality of the identification.  If
SGSGRB's could indeed be detected out to a distance of 100 Mpc
(for fluence thresholds of 10$^{-7}$ ergs cm$^{-2}$), the number
of potential host galaxies would be of the order of several times
$10^4$. With the 1 to 4 arcmin resolution of SWIFT, the
probability of a chance coincidence with any of the 40,000 closest
galaxies is about 0.1 to 1 percent.   In any case, true physical
association could be confirmed by detected afterglow, which we now
consider:

Afterglow from SGR's  can be divided into two classes: a) that
which comes from the neutron star itself or nearby (e.g. Eichler
and Cheng, 1989) and b) that which results in delayed shocks
associated with the expanding fireball (e.g. Paczynski and
Rhoades, 1993). While the latter is now taken for granted in the
context of cosmologically distant GRB's, significant afterglow
depends on the fireball having a significant amount of energy in
some form that creates shocked plasma; pure gamma rays would not
produce significant afterglow. From a very compact region, pairs
in thermal equilibrium mostly disappear relative to photons before
transparency is attained, so it is not clear  {\it a priori} that
there should ever be detectable afterglow from GRB fireballs. The
non-thermal gamma rays, which testify to significant shock energy
release in not-so-compact regions, allowed predictions of
afterglows, but such gamma rays are conspicuously absent from soft
repeaters. (The observed weak radio afterglow following the Aug 27
flare [Frail, Kulkarni and Bloom, 1999] shows that some afterglow
is possible in principle, but there the total gamma ray energy
flux  at Earth was five  orders of magnitude higher than the
fluxes expected in the present scenario.) On the other hand,
non-thermal pairs, baryon contamination, and low frequency
Poynting flux, none of which are sufficiently understood at
present to quantitatively predict or rule out, could all sweep up
interstellar matter, so we will consider afterglow from a possible
plasma fireball as well as from the neutron star.

Consider X-ray afterglow that might be emitted at or near the
surface of the magnetar. The Aug. 27 flare (Woods et al., 2000)
exhibited a power law spectrum, and a peak luminosity in the 2-10
KeV band during the first several hours of order $2 \times 10^{37}
D_5 ^2$erg/s, where $D_5$ is the distance to SGR 1900+14 in units
of 5 Kpc. (n all other cases, an integral subscript n denotes
units of $10^n$ cgs units.) The luminosity subsided as power law
in time roughly as $t^{-0.7}$, remaining above the steady,
pre-flare level for the next 40 days. Thompson and Duncan have
suggested that such emission (as well as the steady emission from
SGR's and AXP's, which typically is also non-thermal) results from
magnetospheric currents induced by the shear imposed by crustal
motion. The electrons and any positrons could Compton scatter
softer emission coming from the surface. The timescale for the
current decay may be established by Compton drag of the softer
photons made at the surface, or the cooling timescale of the crust
itself (Eichler and Cheng, 1989). This observed $t^{-0.7}$ power
law  is consistent with crustal cooling if a) the heat from
magnetic energy release is distributed uniformly throughout the
crust,
 and b) the heat
conductivity and heat capacity are both proportional to
temperature T (Thompson, private communication) as expected in the
zone of the outer crust where the electrons are relativistic and
dominate the heat conduction. Alternatively, the power law decay
could result from dissipation of magnetospheric currents by
Compton cooling or by Ohmic dissipation in shallow layers.
 If one
were to scale up the energetics of the Aug. 27 flare by two orders
of magnitude,  the peak luminosity could rise by at least the same
 factor, and this would yield a total luminosity of up to
 $ 2 \times 10^{39}$ erg/s, as compared to the $2 \times
10^{37}$ erg/s seen following the Aug. 27 event during the first
several hours. This is about the limit for thermal emission below
10 KeV  from a neutron star. It is  just below the level of
detectability of CHANDRA at 100 Mpc, but perhaps could be detected
at somewhat closer distances.

Note that the energy release conjectured here is enough to heat
the entire neutron star interior to $10^9$K. There would then be a
steady X-ray flux of  about $10^{36}$ erg/s lasting thousands of
years. This would constitute  a bright but not atypical anomalous
x-ray pulsar (AXP). That is, if magnetic field energy is released
within magnetars all at once, as opposed to being distributed in
some $10^2$ events  over 10$^4$ years, the SGSGRB contribution to
the AXP population in our own Galaxy and nearby ones would be
comparable to the contribution from SGR activity if the timescale
of the latter is comparable to the timescale of global  neutron
star cooling. If some AXP's have destroyed dipole moments, perhaps
their final field configuration would be expressed in the X-ray
light curves and could be so deciphered with detailed modelling.

 The apparent timescale for late time, optically thin
  afterglow from the expanding
fireball is likely to be of the order of the expansion time of the
fireball divided by $\Gamma^2$.    All other things (including
$\Gamma$) being equal, the timescale would scale as
$E^{\frac{1}{3}}$ (here E is the isotropic equivalent energy in
the non-gamma ray components of the fireball, which can produce
shocks), the luminosity would then scale as $E^{\frac{2}{3}}$, and
the maximum luminosity distance for detectability as
$E^{\frac{1}{3}}$. Since for cosmologically distant GRB's
afterglows are detectable at luminosity distances of order 10 Gpc
at $E_{47} \sim 10^6$, it follows that an afterglow of a fireball
with $E_{47} \sim 1$ might be detectable out to nearly 100 Mpc,
but with a greatly speeded up timescale. Theoretical estimates of
isotropic emission from a mildly or non-relativistic Sedov-von
Neumann- Taylor phase of an expanding blast wave (e.g. Livio and
Waxman, 2000) give a flux of
\begin{eqnarray}
&&f_\nu \approx 1 \left(\frac{1+z}{2}\right)^{1/2}
\left(\frac{\xi_e}{0.3}\right)
 \left(\frac{\xi_B}{0.3}\right)^{3/4}
  n_0^{3/4} E_{51}
d_{28}^{-2} \nu_{\mathrm{GHz}}^{-1/2} \nonumber \\
&&\quad \times \left(\frac{t}{t_{\mathrm{SNT}}}\right)^{-9/10} ~
\mathrm{mJy},
\end{eqnarray}
 where $\xi_e$ and $\xi_B$ are the ratios of fireball
energy that go into relativistic electrons and magnetic field
respectively, $E_{51}$ is the fireball energy in units of
$10^{51}$ ergs, $n_0$ is the number density of the ambient medium
in $cm^{-3}$, $d_{28}$ is the luminosity distance in units of
$10^{28}$cm and t is observer time. This phase is achieved over a
timescale of order $t_{SNT}\sim 10^6(\frac{E}{n_0}^{1/3})s$. This
confirms the above scaling argument that detecting radio afterglow
from SGSGRB's would be difficult but not impossible, depending on
the exact parameters. The
 radio afterglow  flux from 1900+14 ten days after the
Aug. 27 flare [Frail, Kulkarni and Bloom, 1999], for which $E_{51}
d_{28}^{-2} \sim 10^6$,  is about 5 orders of magnitude below what
would be predicted by the above equation with the other
dimensionless parameters chosen to be of order unity, and this
serves as a reminder that afterglows from magnetar events are very
uncertain.

Let us now consider early stage emission from the expanding
fireball, both prompt optical flashes and early afterglow: The
maximum energy E that can emerge (as limited by self-absorption)
from incoherent synchrotron of IC source expanding at bulk Lorentz
factors $\Gamma$ at observed frequency $\nu$ occurring within  an
observed interval $\Delta t$ is given by
\begin{equation}
E \le 0.04 B^{\prime -\frac{1}{7}}[\Gamma \nu \delta t]^3erg
\end{equation}
 (Eichler and Beskin, 2000). Here B$^{\prime}$ is the magnetic field
strength in the fluid frame in Gauss.  The above constraint allows
the luminosity of the optical flash to greatly exceed that of the
host galaxy for the duration of the burst, even if $\delta t \sim
0.1s$ and $\Gamma \sim 10$. It would be a rare, brief event, as is
the case for cosmologically distant GRB's, deserving of searches
by wide angle monitors such as ROTSE II.

It is also worth considering whether there would be a stage of
early afterglow could last at a detectable level long enough to
slew a standard telescope: While the burst is relativistic, the
apparent timescale is speeded up by $\Gamma^2$,  and the energy
released per swept-up hydrogen atom scales as $\Gamma^2$. It can
then be shown that $\Gamma \propto r^{-3/2}$ (Meszaros and Rees,
1997,Waxman, 1997) and, because $t = r/2\Gamma^2 c$, $t \sim
\Gamma^{-8/3} t_{SNT}$. Over timescales of $10^{-8/3} t_{SNT} \sim
2 \times 10^3 [E_{47}/n_o]^{1/3}s$, long enough to slew a standard
telescope, a relativistic shock of $\Gamma \sim 10$ can be picked
up. Inequality (2) admits very large luminosities and implies that
the actual luminosity over $10^3$ s or so is not inhibited at high
$\Gamma$ by self-absorption at optical frequencies but rather by
total energetics and/or proper time elapsed. The total energy
emitted at frequency $\nu$ is given by

\begin{equation}
E_{ph}(\nu)= \epsilon E\xi_e /\Lambda
\end{equation}
where  $\Lambda = \ln(E_{max}/E_{min})\sim 10$ is the number of ln
scales in  energy range of the  accelerated electrons. Here the
optimistic assumption is made that  the spectrum of accelerated
electrons is $E^{-2}$, i.e. with shock energy distributed equally
over electron energy from $E_{min}$ to $E_{max}$.  The efficiency
$\epsilon$
 at which the electrons radiate at observer frequency $\nu$
over an expansion time, is (when not close to unity) given by

\begin{eqnarray}
\epsilon = [\frac{\sigma_T}{4\pi c^2 t^2}] [\frac{\xi_B L t}{m_c
c^2}] [\nu t]^{1/2} [\frac{2L\xi_B
(r_o/c)}{m_ec^2}]^{-1/4}[\frac{1}{2\Gamma^4}]\nonumber\\
  \sim 10^5 \Gamma^{-4}(\xi_B
L_{47})^{3/4}\nu_{15}^{1/2}t_3^{-1/2}
\end{eqnarray}
 Here L is the
energy per unit time passing through a sphere within the shell of
shocked interstellar gas.  Over  observer time $10^3 t_3$s,
assuming it is greater than the actual duration of the gamma ray
burst, the expanding shell of shocked interstellar plasma
increases in thickness to $\sim ct$ so $L_{47}\sim
E_{47}/10^3t_3$.  Using the above equations, we can estimate the
optical afterglow $L_{opt}$ as

\begin{equation}
L_{opt}\sim 10^{41} \xi_e\xi_B^{3/4}E_{47}^{5/4}n_0^{1/2}\nu_{15}
^{1/2}{(\Lambda/10)}^{-1}t_3^{-3/4}erg/s
\end{equation}
The largest uncertainty is in the actual energy of the non-gamma
ray components of the fireball. For optimistic parameters, ($\xi_e
\sim \xi_B \sim 0.3$),  an optical luminosity of about
$10^{40}E_{47}$ erg/s [about mag 26 at 100 Mpc] for $\sim 10^3$s,
might be possible for fireball energies of order $10^{47}$ ergs,
and even this would not easy to detect. Shorter timescales allow
higher luminosities, but demand faster response.
 
In conclusion, we have proposed an instability in magnetars that
may occur well after the SGR phase. It could release much or most
of the magnetic dipole energy  in a single event, and thus produce
an even larger version by a factor of $10^2$ than the largest
flares that have been observed to come from SGR's. (Searches for
nearby GRB's can be motivated by other considerations as well.)
Such events, although less frequent than giant flares from SGR's
by a factor of $f\sim 10^3$, could be seen over a volume of
$f^{3/2}$ times as large as for the giant SGR flares. The prompt
gamma rays and possibly their afterglows  could be observable as
GRB's from nearby galaxies out to a radius of up to, or somewhat
within, 100 Mpc. Apart from directional coincidence with nearby
host galaxies, the afterglows might distinguish this subclass of
bursts, say, by a faster timescale resulting from the much smaller
burst energies than are expected. The contribution of the cooling
crust of the neutron star and/or the persistent currents in the
magnetosphere might contribute to the X-ray afterglow, and, within
perhaps 30 Mpc, the neutron star could perhaps be a CHANDRA source
for as long as several hours. The likelihood of such crustal
afterglow, however, is model dependent, and we hope to report on
the matter in future publications.  Some Galactic AXP's may have
undergone this instability, if the neutron star cooling time is
indeed comparable to that of SGR activity, and would then have
peculiar surface fields.

{\it Note added:} Norris has recently claimed to identify a
separate subclass of long-lag bursts [~0.07 of the total data set]
that correlate with the local supercluster and therefore may be at
a distance of order 50 Mpc. Though the physical motivation is
different in his context, the conclusion  would be consistent with
a burst energy of $10^{47}$ ergs and illustrates, at least, that a
separate subclass of such bursts could in fact be hidden in the
overall data set.

 I thank V. Kaspi, M. Ruderman, Y. Lyubarsky, E. Waxman,
T. Piran, A. Levinson and especially C. Thompson for helpful
discussions. I thank Dr. P.M. Woods {\it et al.} for sharing their
data prior to publication. Much of this work was done while the
author was a guest at the Institute of Theoretical Physics in
Santa Barbara, and supported by NSF grant PHY94 - 07194.
Additional support from the Arnow Chair of Physics and an Adler
Fellowship awarded by the Israel Academy of Sciences is also
acknowledged with gratitude.



\end{document}